%% file: main.tex
\documentclass[10pt,conference]{IEEEtran}
\IEEEoverridecommandlockouts
\usepackage[T1]{fontenc}
\usepackage[utf8]{inputenc}
\usepackage{cite}
\usepackage{amsmath,amssymb}
\usepackage{graphicx}
\usepackage{booktabs,multirow,pifont}
\usepackage{xcolor}
\usepackage{tikz}
\usetikzlibrary{arrows.meta,positioning,fit,calc,decorations.pathreplacing}
\definecolor{blu}{HTML}{2A78D6}\definecolor{ink2}{HTML}{52514E}\definecolor{mut}{HTML}{9A9993}\definecolor{warm}{HTML}{EB6834}
\usepackage{pgfplots}
\pgfplotsset{compat=1.18}
\usepackage{microtype}
\usepackage{url}
\usepackage[hidelinks]{hyperref}
\usepackage[most]{tcolorbox}
\newtcolorbox{rqanswer}{colback=black!6,colframe=black!35,boxrule=0.4pt,arc=2pt,left=3pt,right=3pt,top=1pt,bottom=1pt,before skip=3pt,after skip=3pt}
\newcommand{\sys}{\textsc{Reva}}
\newcommand{\PC}{\mbox{P-C}}
\newcommand{\PV}{\mbox{P-V}}
\newcommand{\PB}{\mbox{P-B}}
\newcommand{\PR}{\mbox{P-R}}
\input{numbers}
\begin{document}

\title{\sys{}: Real-Time Vulnerability Detection through Contextual Probing of Large Language Models}


\author{
    Amine Lbath,
    Massih-Reza Amini,\\
    Université Grenoble Alpes, CNRS, France
}

\maketitle

\begin{abstract}
A vulnerability costs more to fix the later it is discovered in the software lifecycle. As coding agents now generate code faster than developers can review it, software needs a stronger first line of defense that screens each new function in real time, as it is generated. LLM agents have recently become capable of finding real-world vulnerabilities by working as a security expert would: they run analysis tools, explore the surrounding code, test hypotheses and weigh the evidence before forming their own judgement. This reasoning, however, spans many model calls, so that each verdict takes minutes of compute, far too long to keep pace with code generation. Faster AI-based detectors, which fine-tune deep learning models or prompt an LLM for a direct verdict, are considerably less accurate: they judge a function without the context an agent gathers, and an LLM's generated verdict has proven unreliable in prior work even when some of that context is given.

We present \sys{} (REal-time Vulnerability Analysis), an AI vulnerability detector that brings the security context and code understanding components behind agentic detectors into a single forward pass of an LLM. For each new function, \sys{} provides the LLM with a pre-computed threat model of its module, along with lightweight static analysis findings and call-graph information. Rather than asking the LLM for a verdict, \sys{} lets a trained probe read the LLM's internal representations of the function in this context. On PrimeVul's held-out test set, \sys{} on GLM-5.3-Flash reaches \RGFIVETEST\% accuracy in about one second per function. This matches VulTriage (\VTGFIVEACCFAIL\%), the state-of-the-art agentic function-level detector in the literature, run on the same LLM, while being over 250 times faster and less than a tenth as costly. On the same prompt, the probe is 14\% to 17\% more accurate than the LLM's own direct verdict. Across three open LLMs, \sys{} is at least as accurate as VulTriage, and about 10\% more accurate on two of them. These results suggest that an LLM's internal representations already encode much of what agentic detectors obtain through reasoning.
\end{abstract}

\begin{IEEEkeywords}
vulnerability detection, large language models, internal representations, probing, coding agents
\end{IEEEkeywords}

\section{Introduction}

The cost of remediating a vulnerability grows the later it is discovered in the software lifecycle~\cite{boehm2001}. Vulnerabilities are therefore cheapest to fix when they are detected as the code is written. Coding agents now generate code faster than developers can review it, which calls for a first line of defense that screens each new function in real time, as it is generated.

Static analyzers run in real time, but at the scale of agent-generated code, they produce more reports than developers can review, many of them false positives~\cite{johnson2013,du2026fp}. LLM agents have recently become capable of finding real-world vulnerabilities~\cite{cybergym,aixcc} by combining such tools with their own analysis of the code, and agentic detectors bring this approach to function-level detection. VulTriage~\cite{tang2026vultriage}, the state of the art in the function-level detection literature, and VulAgent~\cite{wang2026vulagent} enrich the prompt with retrieved vulnerability knowledge and program facts, and VulTrial~\cite{widyasari2026vultrial} lets several agents debate a verdict. This reasoning, however, spans several model calls, which amounts to minutes per function. Faster AI-based detectors, which fine-tune deep learning models or prompt an LLM for a direct verdict, reduce this latency but are considerably less accurate~\cite{ding2025primevul,wen2025revd,ovadia2026mist}.

\input{fig_method}

\sys{} is designed to close this gap between speed and improved accuracy. Instead of letting the LLM gather context through multiple rounds of tool calls, \sys{} assembles it in the prompt, including a threat model of the function's module, its static analysis findings, its call graph and its file. Instead of reasoning towards a verdict, the LLM then processes this prompt in a single forward pass without generating text, and a small probe trained on labeled functions maps its internal representations of the code tokens to a risk score. This second choice builds on evidence that LLMs encode knowledge that their answers do not express, for example about the truth of a statement~\cite{azaria2023,burns2023,marks2024} or the correctness of their own answers~\cite{orgad2025,kadavath2022}, and that chain-of-thought explanations are not always faithful to the underlying computation~\cite{turpin2023}.

We evaluate \sys{} on PrimeVul~\cite{ding2025primevul}, a high-quality dataset of vulnerable C and C++ functions paired with benign versions of the same functions, which spans \PVPROJECTS{} projects and \PVCWES{} vulnerability types (CWEs). On its held-out test set, \sys{} on GLM-5.3-Flash~\cite{glm53f} is on par with VulTriage and more accurate than VulAgent run on the same LLM, in about one second per function, over 250 times faster than VulTriage. On two other open LLMs, it is 9\% to 13\% more accurate than both agentic detectors.

This paper makes three contributions: (i) we propose \sys{}, a vulnerability detector that combines targeted security context with a probe on an LLM's internal representations, in a single forward pass per function; (ii) we compare it with state-of-the-art agentic function-level detectors run on the same three LLMs, on PrimeVul's held-out test set, in terms of accuracy, cost and robustness; and (iii) we outline a research agenda towards detectors that screen every function as coding agents generate it and escalate likely vulnerabilities to more capable agents that construct a proof of vulnerability.

\section{Background and Related Work}

\textbf{LLM-based vulnerability detection.} LLMs in non-agentic settings typically struggle to tell a vulnerable function from its benign counterpart: on PrimeVul~\cite{ding2025primevul}, GPT-4 with chain-of-thought labels both versions correctly in only \GPTFOURCOTPC\% of the pairs, and LLMs asked for a verdict in non-agentic settings, often reason inconsistently about security~\cite{ullah2024,steenhoek2025}. \sys{} instead reads the LLM's internal representations rather than its answer or reasoning.

\textbf{Agentic detectors.} Recent detectors improve accuracy by adding knowledge and reasoning steps. VulTriage retrieves CWE descriptions and verbalizes control- and data-flow facts~\cite{tang2026vultriage}, VulAgent lets specialized agents propose and validate hypotheses against the project's code~\cite{wang2026vulagent}, and VulTrial stages a debate between agents~\cite{widyasari2026vultrial}. Their cost grows with each call, and their context can exceed the LLM's context window. \sys{} instead pre-loads the key parts of this context into the prompt and replaces the multi-step reasoning with a single probed forward pass.

\textbf{Probing LLM internal representations.} Probes are small classifiers trained on a model's hidden states~\cite{belinkov2022}. For code, linear probes indicates some capability to separate secure from insecure code on toy datasets~\cite{yu2025moc, khatri2026prefill}, and predict whether generated code will be insecure or incorrect~\cite{sternfeld2026fragility,dicicco2026decodable}. These probes read the code alone, whereas \sys{} probes real-world functions read with targeted security context.

\textbf{Pitfalls of vulnerability benchmarks.} Detectors can exploit spurious features of their datasets, such as code metrics or semantics unrelated to the vulnerability~\cite{risse2024,weissberg2026codemetrics}. Such a detector tends to classify a vulnerable function and its benign counterpart with the same label, which share these features. We therefore score each function independently, then check whether the two versions of each pair receive their correct labels.

\section{Contextual Probing with \sys{}}

In \sys{}, an LLM reads a new function with its security context in one forward pass, and a probe maps the hidden states of the code tokens to a risk score (Fig.~\ref{fig:pipeline}).

\textbf{\ding{172} Prompt with security context.} The prompt contains the new function and the context around it, in labeled sections (Fig.~\ref{fig:pipeline}). A threat model, built once per project by a workflow of LLM agents over the entire software, describes the project (summary, attacker-controlled inputs, trust boundaries) and the module that contains the function (role, inputs that reach it, typical sinks and guards). The prompt adds the Semgrep~\cite{semgrep} findings for the function, its call sites and the signatures of the helpers it calls, and the includes, types, macros and sibling functions of its file, extracted with tree-sitter from the repository. It is available for 84\% of the training functions; otherwise the prompt has the function and its Semgrep findings only. The median prompt has \RGLIMPROMPTTOK{} tokens, \RGLIMCODETOK{} of them code.

\textbf{\ding{173} One forward pass.} The LLM processes the prompt in one forward pass that stops after the deepest of four probed layers, so the remaining layers are never run and it generates no output token. We probe intermediate layers, which probing studies often find most informative~\cite{belinkov2022}: four evenly spaced layers from 38\% to 85\% of the depth of the LLM, selected by cross-validation.

\textbf{\ding{174} Code-token states.} At each probed layer, \sys{} keeps only the hidden states of the function's tokens. Each state is normalized and projected to \PROJDIM{} dimensions, and the four layers are concatenated.

\textbf{\ding{175} Attention probe.} A probe with \NHEADS{} attention heads pools the token states into one risk score. It has \PROBEPARAMS{} parameters and is trained on PrimeVul's training functions. Its loss combines a per-function classification term with a term that ranks each vulnerable function above its benign counterpart. These pairs serve only in training: \sys{} scores each function on its own.

\textbf{Use in practice.} The LLM and Semgrep's rules are loaded once at startup, so they don't have to be booted for each new analysis, and each project's threat model is built ahead of time. For each new function, \sys{} extracts its context, runs Semgrep, builds the prompt, runs the pass and applies the probe. With \GLIMNAME{} on one H200 GPU, this takes \RGLIMEE\,s per function, \CTXSEC\,s of which is context extraction.

\section{Preliminary Results}
\label{sec:results}

We evaluate \sys{} through three research questions:
\begin{itemize}\setlength{\itemsep}{0pt}\setlength{\topsep}{2pt}\setlength{\parsep}{0pt}
\item[] \textbf{RQ1:} How accurate is \sys{} compared with agentic detectors running on the same LLM?
\item[] \textbf{RQ2:} How much computation does \sys{} save?
\item[] \textbf{RQ3:} Which parts of \sys{} matter, and is its accuracy robust to leakage and selection?
\end{itemize}

\begin{figure}[t]
\centering
\includegraphics[width=\columnwidth]{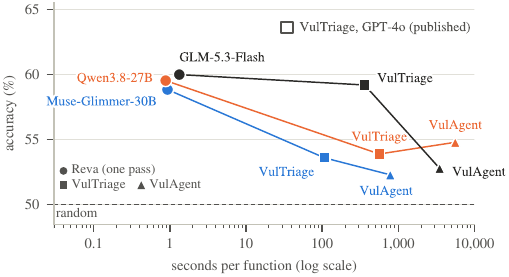}
\caption{Accuracy versus time per function on the PrimeVul test set, one color per LLM. Functions without a verdict count as benign. Hollow marker: VulTriage on GPT-4o, as reported.}
\label{fig:trend}
\end{figure}

\subsection{Experimental Setup}

\textbf{Dataset and metrics.} PrimeVul~\cite{ding2025primevul} pairs each vulnerable C/C++ function with a benign version of the same function. It has \NTRAINPAIRS{} training, \NVALIDPAIRS{} validation and \NTESTPAIRS{} test pairs. Following \cite{ding2025primevul}, we count the pairs in which both versions are labeled correctly (\PC), both are flagged (\PV), both are cleared (\PB), or the labels are reversed (\PR). Our main metric is accuracy over all functions, each scored independently.

\textbf{Protocol.} We made every design choice on the training and validation sets only. Cross-validation (CV) accuracy covers five folds of the training set, with all functions of a commit in the same fold, and the validation set, scored by a probe trained on the full training set. The test set stayed unseen during development and was scored only with the final configuration.

\textbf{LLMs.} We use three open-weight LLMs: \GLIMNAME{}~\cite{glimmer30b}, Qwen3.8-27B~\cite{qwen38} and GLM-5.3-Flash~\cite{glm53f}, a larger mixture-of-experts model. On GLM-5.3-Flash, the agentic detectors use the FP8 version of the LLM and \sys{} an exact BF16 conversion, which leaves its cross-validation accuracy unchanged (\BFDEV\% vs.\ \RGFIVEDEVPOINT\%).

\textbf{Baselines.} We run VulAgent and VulTriage on their released code, using the same LLMs as \sys{}. A function is \emph{failed} when the detector returns no verdict, usually because the reasoning exceeds the token limit. We compare all methods over all functions, counting failed functions as benign. Published GPT-4o results are given for context.

\begin{table*}[t]
\caption{Accuracy and cost of \sys{} and agentic detectors on the PrimeVul test set. Tokens: mean input and output tokens per function, in thousands. Time: median seconds per function. CI: pair-bootstrap interval on accuracy. Cost: USD per 1{,}000 functions, for \sys{} on one rented H200 GPU at \$3.50 per hour, and for the agentic detectors at public API rates per million input/output tokens (\GLIMNAME{} \$0.35/\$1.50, Qwen3.8-27B \$0.40/\$3.00, GLM-5.3-Flash \$0.15/\$0.50). FPR: share of non-vulnerable functions flagged. --: not reported.}
\label{tab:main}
\centering\footnotesize
\setlength{\tabcolsep}{5pt}
\begin{tabular}{lcrrrccccc}
\toprule
Method and LLM & Calls & Tokens (k) & Time (s) & Cost (\$) & FPR & Accuracy & 95\% CI & \PC & \PR\\
\midrule
\multicolumn{10}{l}{\textbf{\sys{}} (ours)} \\
\quad \GLIMNAME{} & 1 & \RGLIMTOK{} / 0 & \RGLIMEE & 0.9 & \RGLIMTESTFPR & \RGLIMTEST & [56.8, 61.0] & \RGLIMTESTPC & \RGLIMTESTPR\\
\quad Qwen3.8-27B & 1 & \RQWENTOK{} / 0 & \RQWENEE & 0.9 & \RQWENTESTFPR & \RQWENTEST & [57.4, 61.7] & \RQWENTESTPC & \RQWENTESTPR\\
\quad GLM-5.3-Flash & 1 & \RGFIVETOK{} / 0 & \RGFIVEEE & 1.3 & \RGFIVETESTFPR & \textbf{\RGFIVETEST} & [57.8, 62.2] & \RGFIVETESTPC & \RGFIVETESTPR\\
\quad \GLIMNAME{} + GLM-5.3-Flash & 2 & 6.3 / 0 & \RSTACKEE & 1.7 & \RSTACKTESTFPR & \textbf{\RSTACKTEST} & [58.5, 62.9] & \RSTACKTESTPC & \RSTACKTESTPR\\
\midrule
\multicolumn{10}{l}{\textbf{VulTriage}~\cite{tang2026vultriage}} \\
\quad \GLIMNAME{} & 3 & \VTGLIMTESTIN{} / \VTGLIMTESTOUT & \VTGLIMTESTSEC & 9.4 & \VTGLIMTESTFPR & \VTGLIMTESTACCFAIL & [51.5, 55.7] & \VTGLIMTESTPC & \VTGLIMTESTPR\\
\quad Qwen3.8-27B & \VTQWENCALLS & \VTQWENIN{} / \VTQWENOUT & \VTQWENSEC & 61.4 & \VTQWENFPR & \VTQWENACCFAIL & [51.6, 56.2] & \VTQWENPC & \VTQWENPR\\
\quad GLM-5.3-Flash & \VTGFIVECALLS & \VTGFIVEIN{} / \VTGFIVEOUT & \VTGFIVESEC & 13.3 & \VTGFIVEFPR & \VTGFIVEACCFAIL & [56.7, 61.6] & \VTGFIVEPC & \VTGFIVEPR\\
\quad \textcolor{black!50}{GPT-4o (published, not reproducible)} & \textcolor{black!50}{3} & \textcolor{black!50}{--} & \textcolor{black!50}{\VTGPTSEC} & \textcolor{black!50}{--} & \textcolor{black!50}{\VTGPTFPR} & \textcolor{black!50}{\VTGPTACC} & \textcolor{black!50}{--} & \textcolor{black!50}{\VTGPTPC} & \textcolor{black!50}{\VTGPTPR}\\
\midrule
\multicolumn{10}{l}{\textbf{VulAgent}~\cite{wang2026vulagent}} \\
\quad \GLIMNAME{} & \VAGLIMCALLS & \VAGLIMIN{} / \VAGLIMOUT & \VAGLIMSEC & 62.6 & \VAGLIMFPR & \VAGLIMACC & [50.0, 54.4] & \VAGLIMPC & \VAGLIMPR\\
\quad Qwen3.8-27B & \VAQWENCALLS & \VAQWENIN{} / \VAQWENOUT & \VAQWENSEC & 306.9 & \VAQWENFPR & \VAQWENACCFAIL & [52.0, 57.0] & \VAQWENPC & \VAQWENPR\\
\quad GLM-5.3-Flash & 9.3 & \VASIXIN{} / \VASIXOUT & \VASIXSEC & 97.8 & \VAGFIVEFPRALL & \VAGFIVEACCALL & [50.5, 54.9] & \VAGFIVEPCALL & \VAGFIVEPRALL\\
\quad \textcolor{black!50}{GPT-4o (published)} & \textcolor{black!50}{\VAGPTCALLS} & \textcolor{black!50}{-- / \VAGPTOUT} & \textcolor{black!50}{--} & \textcolor{black!50}{--} & \textcolor{black!50}{\VAGPTFPR} & \textcolor{black!50}{\VAGPTACC} & \textcolor{black!50}{--} & \textcolor{black!50}{\VAGPTPC} & \textcolor{black!50}{\VAGPTPR}\\
\midrule
\multicolumn{10}{l}{\textbf{VulTrial}~\cite{widyasari2026vultrial}} \\
\quad \textcolor{black!50}{GPT-4o (published)} & \textcolor{black!50}{--} & \textcolor{black!50}{--} & \textcolor{black!50}{--} & \textcolor{black!50}{--} & \textcolor{black!50}{\VTRIALFPR} & \textcolor{black!50}{\VTRIALACC} & \textcolor{black!50}{--} & \textcolor{black!50}{\VTRIALPC} & \textcolor{black!50}{\VTRIALPR}\\
\bottomrule
\end{tabular}
\\[2pt]\raggedright\scriptsize
\end{table*}

\begin{figure}[t]
\centering
\includegraphics[width=0.9\columnwidth]{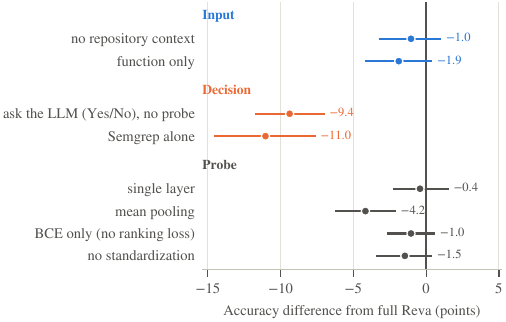}
\caption{Change in validation accuracy of \sys{} on \GLIMNAME{} when one component is removed or replaced, with paired 95\% confidence intervals.\ABLCAPTBD}
\label{fig:ablation}
\end{figure}

\subsection{RQ1: Accuracy}

On GLM-5.3-Flash, \sys{} reaches \RGFIVETEST\% accuracy (95\% CI \RGFIVETESTCI), against \VTGFIVEACCFAIL\% for VulTriage and \VAGFIVEACCALL\% for VulAgent run on the same LLM (Table~\ref{tab:main}). Over all test functions, \sys{} is \DAVT{} points above VulTriage (CI \DAVTCI) and \DAVA{} points above VulAgent (CI \DAVACI). On Qwen3.8-27B and \GLIMNAME{}, \sys{} is 5.1 to 6.7 points more accurate than both agentic detectors, and all four confidence intervals exclude zero.

Beyond accuracy, \sys{} also compares favorably with VulTriage, the state-of-the-art function-level agentic detector, when both run on the same LLM. On all three LLMs, \sys{} has a lower false positive rate (FPR) and inverts the labels of fewer pairs (\PR), and on two of them it labels more pairs entirely correctly (\PC). \sys{} thus not only matches or exceeds VulTriage in accuracy, but also separates vulnerable from benign functions more reliably, which suggests that the probe captures the vulnerability itself rather than surface features that a vulnerable function shares with its secured counterpart.

Fig.~\ref{fig:trend} summarizes this trade-off. On each LLM, \sys{} is at least as accurate as both agentic detectors while being 100 to 6{,}000 times faster. Its accuracy also varies little across LLMs, by 1.1 points against 5.6 for VulTriage, so that \sys{} on \GLIMNAME{} matches VulTriage on GLM-5.3-Flash, the LLM on which VulTriage performs best. The signal read by the probe is therefore present in all three LLMs, whereas the accuracy of the agentic detectors depends on how well each LLM reasons about the code. VulTriage on GPT-4o reports a higher accuracy (\VTGPTACC\%), but we were unable to replicate these results.

\begin{rqanswer}\emph{Answer to RQ1:} From a single forward pass, \sys{} matches or exceeds both function-level agentic detectors on all three LLMs. It is significantly more accurate in five of the six same-LLM comparisons and has fewer false positives than VulTriage on every LLM.\end{rqanswer}

\subsection{RQ2: Computational Cost}

End to end, \sys{} takes \RQWENEE\,s to \RGFIVEEE\,s per function, against \VTGLIMTESTSEC\,s to \VAQWENSEC\,s for the agentic detectors on the same LLMs. Although these times, measured under parallel load, overstate single-request latency, the token counts confirm the gap. \sys{} reads about 3{,}100 to 3{,}400 tokens per function and generates none, whereas the agentic detectors process \TOKRATIOLO{} to \TOKRATIOHI{} times more tokens, many of them generated. As \sys{} needs the LLM's hidden states, it cannot run behind an API. On a rented H200 GPU, it screens 1{,}000 functions for \$0.9 to \$1.3, against \$9 to \$61 for VulTriage and \$63 to \$307 for VulAgent at the public API rates of the same LLMs.

\begin{rqanswer}\emph{Answer to RQ2:} \sys{} screens a function in about one second, against minutes for the agentic detectors, at 10 to 360 times lower cost on the same LLM, and returns a score for every function.\end{rqanswer}

\subsection{RQ3: Ablation and Robustness}
\label{sec:rq3}

We ablate each component of \sys{} on the \GLIMNAME{} validation set (Fig.~\ref{fig:ablation}). Asking the same LLM for a Yes/No verdict on the same prompt, instead of probing it, reduces accuracy to chance. On the test set, it stays \DASKQWEN{} to \DASKGLIM{} points below the probe, and Semgrep, Cppcheck and Flawfinder alone are also at chance (\SGALONETEST\% to \FLAWFINDERTEST\%). The LLM's internal representations thus carry information that neither its verdict nor the static-analysis findings provide alone. Within the probe, replacing attention pooling by a plain average of the token states lowers accuracy by 4.2 points, which suggests that the evidence of a vulnerability lies in a few tokens. Using a single layer, omitting the standardization or dropping the ranking term lowers accuracy by 0.4 to 1.5 points, and removing the repository context or keeping only the function by 1.0 and 1.9 points, none of which is significant. On PrimeVul's short functions, the function alone thus carries most of the signal; we expect the context to matter more for interprocedural vulnerabilities.

Leakage and selection on the evaluation data can inflate probing results~\cite{belinkov2022,hewitt2019}. When no project appears in both training and evaluation folds, cross-validation accuracy drops by only 1.3 to 2.7 points, so most of what the probe learns transfers to unseen projects. Likewise, choosing the four probed layers on the training folds only, instead of on all development data, changes accuracy by less than half a point, so the layer choice does not inflate the results. Accuracy is, however, 6 points lower on the test set than in cross-validation (\RGFIVETEST\% against \BFDEV\% for GLM-5.3-Flash). As PrimeVul splits its data chronologically, this gap likely reflects the shift towards newer code, which cross-validation does not capture.

\begin{rqanswer}\emph{Answer to RQ3:} The probe is essential. A direct verdict from the same LLM falls to chance, and replacing attention pooling by an average of the token states costs 4.2 points, whereas the repository context has no significant effect on PrimeVul. Accuracy largely holds on unseen projects.\end{rqanswer}

\section{Future Plans}
\label{sec:future}

\textbf{Step 1: Complete the comparison.} We will run both agentic detectors and \sys{} on larger open LLMs. We will also test whether reading the internal representations after a short self-generated analysis adds information. \emph{Success criterion:} \sys{} matches or exceeds them on the same LLM at every scale.

\textbf{Step 2: Calibrated screening in coding agents.} We will learn a project- and function-dependent threshold, and integrate \sys{} into a coding-agent loop that escalates only high-risk functions to an agentic detector or a proof-of-vulnerability check~\cite{cyberforge}. \emph{Success criterion:} for a given budget of escalations, the loop finds more vulnerabilities than the agentic detector alone.

\textbf{Step 3: Generalization beyond PrimeVul.} We will evaluate on other datasets~\cite{sven,diversevul,aviator}, on interprocedural vulnerabilities~\cite{lbath2026repo,cyberforge} and on vulnerabilities disclosed after each LLM's training cutoff, and study fine-tuning the LLM with the same objective. \emph{Success criterion:} \sys{} keeps its accuracy on vulnerabilities the LLM cannot have seen.

\section{Conclusion}

\sys{} detects vulnerabilities by probing an LLM's internal representations of a function and its security context in a single forward pass. On PrimeVul's held-out test set, it matches or exceeds agentic detectors run on the same LLM, with fewer false positives, in about one second per function and at a small fraction of their cost. An LLM's internal representations thus carry more information about vulnerabilities than its verdicts, and reading them enables real-time screening of every function as it is written.

\clearpage
\bibliographystyle{IEEEtran}
\bibliography{refs}

\end{document}

%% file: numbers.tex
\newif\ifrowA \ifdefined\ROWFALLBACK \rowAfalse \else \rowAtrue \fi
\newif\ifunsealed \unsealedtrue   

\newcommand{\GLIMNAME}{Muse-Glimmer-30B}                      

\newcommand{\NTRAINPAIRS}{3{,}789}
\newcommand{\NVALIDPAIRS}{480}

\newcommand{\NTESTPAIRS}{435}

\newcommand{\PROJDIM}{1{,}024}
\newcommand{\NHEADS}{8}
\newcommand{\PROBEPARAMS}{0.57\,M}

\newcommand{\RGLIMPROMPTTOK}{3{,}840}
\newcommand{\RGLIMCODETOK}{601}

\newcommand{\RGLIMTEST}{58.9}          
\newcommand{\RGLIMTESTPC}{20.2}        
\newcommand{\RGLIMTESTPR}{2.5}\newcommand{\RGLIMTESTFPR}{41.1}

\newcommand{\RQWENTEST}{59.5}           
\newcommand{\RQWENTESTPC}{22.1}

\newcommand{\RQWENTESTPR}{3.0}\newcommand{\RQWENTESTFPR}{40.5}

\newcommand{\VTGLIMTESTACCFAIL}{53.6}\newcommand{\VTGLIMTESTPC}{14.0}\newcommand{\VTGLIMTESTPR}{6.9}
\newcommand{\VTGLIMTESTFPR}{75.6}\newcommand{\VTGLIMTESTSEC}{108}\newcommand{\VTGLIMTESTIN}{6.2}\newcommand{\VTGLIMTESTOUT}{4.8}

\newcommand{\RGFIVEDEVPOINT}{65.9}

\newcommand{\RGFIVETEST}{60.0}          
\newcommand{\RGFIVETESTPC}{22.3}

\newcommand{\RGFIVETESTPR}{2.3}
\newcommand{\RGFIVETESTFPR}{40.0}
\newcommand{\RGFIVETESTCI}{[57.8, 62.2]}

\newcommand{\RSTACKTEST}{60.7}         
\newcommand{\RSTACKTESTPC}{23.2}
\newcommand{\RSTACKTESTPR}{1.8}
\newcommand{\RSTACKTESTFPR}{39.3}

\newcommand{\VTQWENACCFAIL}{53.9}
\newcommand{\VTQWENFPR}{75.4}

\newcommand{\VTQWENCALLS}{3}
\newcommand{\VTQWENIN}{8.0}           
\newcommand{\VTQWENOUT}{19.4}
\newcommand{\VTQWENSEC}{564}           
\newcommand{\VTQWENPC}{16.1}

\newcommand{\VTQWENPR}{8.3}

\newcommand{\VAQWENACCFAIL}{54.5}
\newcommand{\VAQWENFPR}{32.2}

\newcommand{\VAQWENCALLS}{9.2}
\newcommand{\VAQWENIN}{37.4}
\newcommand{\VAQWENOUT}{97.3}
\newcommand{\VAQWENSEC}{5{,}614}
\newcommand{\VAQWENPC}{20.2}
\newcommand{\VAQWENPR}{11.3}

\newcommand{\VTGFIVEACCFAIL}{59.2}
\newcommand{\VTGFIVEFPR}{46.0}

\newcommand{\VTGFIVECALLS}{3}
\newcommand{\VTGFIVEIN}{6.2}
\newcommand{\VTGFIVEOUT}{24.7}
\newcommand{\VTGFIVESEC}{362}
\newcommand{\VTGFIVEPC}{26.0}

\newcommand{\VTGFIVEPR}{7.6}

\newcommand{\VAGLIMACC}{52.2}
\newcommand{\VAGLIMFPR}{64.1}
\newcommand{\VAGLIMPC}{13.3}
\newcommand{\VAGLIMPR}{9.0}
\newcommand{\VAGLIMSEC}{785}
\newcommand{\VAGLIMIN}{34.1}
\newcommand{\VAGLIMOUT}{33.8}
\newcommand{\VAGLIMCALLS}{9.2}

\newcommand{\VTGPTACC}{63.6}
\newcommand{\VTGPTPC}{33.8}
\newcommand{\VTGPTFPR}{54.3}
\newcommand{\VTGPTSEC}{34.5}
\newcommand{\VAGPTACC}{58.4}
\newcommand{\VAGPTPC}{26.6}
\newcommand{\VAGPTFPR}{36.7}
\newcommand{\VAGPTCALLS}{8.8}
\newcommand{\VAGPTOUT}{7.0}
\newcommand{\VTRIALACC}{53.6}         
\newcommand{\VTRIALPC}{18.6}
\newcommand{\VTRIALFPR}{52.6}
\newcommand{\VTRIALPR}{11.5}        
\newcommand{\VAGPTPR}{10.1}
\newcommand{\VTGPTPR}{6.7}          

\newcommand{\GPTFOURCOTPC}{12.9}

\newcommand{\ABLCAPTBD}{}

\newcommand{\DASKGLIM}{8.7}\newcommand{\DASKQWEN}{7.4}
\newcommand{\SGALONETEST}{50.8}
\newcommand{\FLAWFINDERTEST}{51.3}   

\newcommand{\TOKRATIOLO}{3}      
\newcommand{\TOKRATIOHI}{70}      
\newcommand{\RGLIMTOK}{3.2}\newcommand{\RQWENTOK}{3.4}\newcommand{\RGFIVETOK}{3.1}

\newcommand{\DAVT}{0.8}    \newcommand{\DAVTCI}{[$-$2.1, +3.8]}       
\newcommand{\DAVA}{7.2}    \newcommand{\DAVACI}{[+4.4, +10.3]}       
\newcommand{\VAGFIVEACCALL}{52.8}   

\newcommand{\VASIXIN}{35.8}\newcommand{\VASIXOUT}{184.8}
\newcommand{\VASIXSEC}{3{,}495}   
\newcommand{\BFDEV}{66.2}

\newcommand{\CTXSEC}{0.46}\newcommand{\RGLIMEE}{0.93}\newcommand{\RQWENEE}{0.88}\newcommand{\RGFIVEEE}{1.33}\newcommand{\RSTACKEE}{1.78}
\newcommand{\VAGFIVEPCALL}{14.3}\newcommand{\VAGFIVEPRALL}{8.7}\newcommand{\VAGFIVEFPRALL}{19.8}
\newcommand{\PVPROJECTS}{651}\newcommand{\PVCWES}{121}

%% file: fig_method.tex
\begin{figure*}[t]
\centering
\begin{tikzpicture}[font=\scriptsize,
  panel/.style={draw=ink2!70,dashed,rounded corners=3pt},
  ptitle/.style={font=\scriptsize\bfseries,text=ink2,anchor=north west},
  lbl/.style={align=center,text=mut,font=\fontsize{6.5}{7.5}\selectfont},
  box/.style={draw=ink2,rounded corners=1.5pt,align=center,inner sep=2pt,fill=white,font=\scriptsize},
  arr/.style={-{Stealth[length=3.5pt]},ink2,semithick},
  card/.style={draw=ink2!60,fill=white,rounded corners=1.5pt},
  layer/.style={draw=blu!80!black,rounded corners=1pt,minimum width=1.5cm,minimum height=0.15cm,inner sep=0pt}]

\draw[panel,fill=black!2] (0,0) rectangle (5.4,3.2);
\node[ptitle] at (0.1,3.13) {\ding{172}\ Prompt with security context};
\draw[card,draw=ink2,fill=white] (0.15,0.46) rectangle (5.25,2.76);
\begin{scope}[every node/.style={anchor=west,inner sep=0pt,font=\fontsize{6}{7}\selectfont\ttfamily,text=blu!80!black}]
  \node at (0.27,2.62) {\#\# Threat model: project, module};
  \node at (0.27,2.23) {\#\# Semgrep findings};
  \node at (0.27,1.96) {\#\# Call graph: callers, callees};
  \node at (0.27,1.58) {\#\# File: includes, types, macros};
  \node at (0.27,1.25) {\#\# New function};
\end{scope}
\begin{scope}[mut!45,line width=1.6pt,line cap=round]
  \draw (0.3,2.48) -- (4.6,2.48); \draw (0.3,2.38) -- (3.7,2.38);
  \draw (0.3,2.1) -- (3.2,2.1);
  \draw (0.3,1.82) -- (4.3,1.82); \draw (0.3,1.72) -- (2.9,1.72);
  \draw (0.3,1.45) -- (4.8,1.45); \draw (0.3,1.35) -- (3.9,1.35);
\end{scope}
\begin{scope}[every node/.style={anchor=west,inner sep=0pt,font=\fontsize{6}{7}\selectfont\ttfamily,text=ink2}]
  \node at (0.3,1.03) {node *new\_node(tree *tr) \{};
  \node at (0.3,0.81) {\ node *t = \&tr->nodes[tr->n];};
  \node at (0.3,0.59) {\ tr->n++; return t; \}};
\end{scope}
\node[lbl] at (2.7,0.21) {security context around the new function};

\draw[panel,fill=blu!3] (5.6,0) rectangle (9.1,3.2);
\node[ptitle] at (5.7,3.13) {\ding{173}\ One forward pass};
\foreach \i in {0,...,8} {
  \pgfmathtruncatemacro{\f}{8+5*\i}
  \node[layer,fill=blu!\f] (L\i) at (6.75,0.62+0.19*\i) {};
}
\foreach \i in {3,5,7,8} { \node[layer,draw=blu!85!black,line width=0.9pt,fill=blu!55] at (6.75,0.62+0.19*\i) {};
  \draw[-{Stealth[length=2.5pt]},blu!85!black,semithick] (7.52,0.62+0.19*\i) -- (8.2,0.62+0.19*\i); }
\foreach \i in {9,10} { \node[layer,draw=mut,dashed,fill=white] at (6.75,0.62+0.19*\i) {}; }
\node[lbl,anchor=west,align=left] at (7.62,2.4) {not\\executed};
\node[lbl,anchor=west,text=blu!85!black] at (8.25,1.05) {probed};
\node[lbl] at (7.35,0.25) {LLM, no text generated};

\draw[panel,fill=black!2] (9.3,0) rectangle (12.9,3.2);
\node[ptitle] at (9.4,3.13) {\ding{174}\ Code-token states};
\foreach \g in {0,...,3} {
  \pgfmathsetmacro{\gy}{0.8+0.5*\g}
  \foreach \r in {0,1,2} { \foreach \c in {0,...,11} {
    \pgfmathtruncatemacro{\v}{mod(7*\c*\c+13*\r+5*\c*\r+11*\g+3,58)+6}
    \fill[blu!\v] (10.05+0.18*\c,\gy+0.15*\r) rectangle ++(0.18,0.15);
  }}
  \draw[white,line width=0.3pt] (10.05,\gy) grid[xstep=0.18,ystep=0.15] (12.21,\gy+0.45);
  \draw[ink2,line width=0.4pt] (9.95,\gy) rectangle (12.21,\gy+0.45);
  \pgfmathtruncatemacro{\gl}{\g+1}
  \node[anchor=east,text=ink2,font=\fontsize{6.5}{7.5}\selectfont] at (10.0,\gy+0.225) {$\ell_{\gl}$};
}
\draw[ink2,decorate,decoration={brace,amplitude=3pt}] (12.28,2.75) -- (12.28,0.8) node[midway,right=2pt,text=ink2,font=\fontsize{6.5}{7.5}\selectfont,align=left] {con-\\cat};
\node[lbl] at (11.1,0.25) {$T$ code tokens $\times$ 4$\cdot$1{,}024};

\draw[panel,fill=blu!3] (13.1,0) rectangle (17.9,3.2);
\node[ptitle] at (13.2,3.13) {\ding{175}\ Attention probe $\to$ risk score};
\foreach \k/\h in {0/0.08,1/0.12,2/0.10,3/0.18,4/0.62,5/0.70,6/0.22,7/0.10,8/0.07,9/0.12,10/0.09,11/0.06,12/0.08,13/0.05} {
  \ifnum\k=4 \def\bc{warm}\else\ifnum\k=5 \def\bc{warm}\else\def\bc{blu!55}\fi\fi
  \fill[\bc] (13.55+0.27*\k,1.9) rectangle ++(0.17,\h);
}
\draw[ink2,line width=0.4pt] (13.45,1.9) -- (17.4,1.9);
\node[lbl] at (15.45,1.72) {learned attention over code tokens};
\shade[left color=blu!12,right color=warm!55] (13.5,1.0) rectangle (17.5,1.14);
\draw[ink2,line width=0.4pt] (13.5,1.0) rectangle (17.5,1.14);
\draw[ink2,line width=0.8pt] (15.5,0.92) -- (15.5,1.22);
\node[lbl,anchor=north] at (15.5,0.92) {median threshold};
\fill[warm] (16.7,1.2) -- ++(-0.08,0.13) -- ++(0.16,0) -- cycle;
\node[anchor=south,inner sep=1pt,text=warm!85!black,font=\fontsize{6.5}{7.5}\selectfont] at (16.7,1.33) {function's risk score};
\node[lbl] at (15.5,0.25) {probe of \PROBEPARAMS{} parameters, trained on labels};

\draw[arr] (5.4,1.6) -- (5.6,1.6);
\draw[arr] (8.55,1.6) -- (9.3,1.6);
\draw[arr] (12.9,1.6) -- (13.1,1.6);
\end{tikzpicture}
\caption{Overview of \sys{} on a vulnerable function from LibRaw (CVE-2020-35530). \ding{172} The prompt combines the function with its security context. \ding{173} The LLM processes the prompt in one forward pass, up to the deepest of four probed layers. \ding{174} The hidden states of the code tokens are kept at each probed layer. \ding{175} An attention probe pools them into a risk score.}
\label{fig:pipeline}
\end{figure*}